\documentclass[pre,twocolumn,twoside,showpacs,superscriptaddress]{revtex4}
\usepackage{graphics,amssymb,amsmath,epsf,color,hyperref}
\epsfclipon

\begin{document}
\title{Winding number excitation detects phase transition in one-dimensional $XY$ model with variable interaction range}
\author{Hyunsuk Hong}
\affiliation{Department of Physics and Research Institute of Physics
and Chemistry, Chonbuk National University, Jeonju 561-756, Republic
of Korea}
\author{Beom Jun Kim}
\email[Corresponding author: ]{beomjun@skku.edu}
\affiliation{Department of Physics, Sungkyunkwan University, Suwon 440-746,
Republic of Korea} 
\date{\today}

\begin{abstract}
We numerically study the critical behavior of the
one-dimensional $XY$ model of the size $N$ with variable interaction range $L$.
As expected, the standard local order parameter of the magnetization is shown
to well detect the mean-field type transition which occurs 
at any nonzero value of $L/N$. 
The system is particularly interesting since the underlying one-dimensional 
structure allows us to study the topological excitation of the winding number 
across the whole system even though the system shares the 
mean-field transition with the globally-coupled system.  
We propose a novel nonlocal order parameter
based on the width of the winding number distribution
which exhibits a clear signature of the transition nature of the system. 
\end{abstract}
\maketitle

\section{Introduction}
The $XY$ model is one of typical model systems
that have been widely studied in the arena of 
statistical physics~\cite{ref:XYreview}.
Most existing studies so far have been performed in the regular lattice
structures of one~\cite{ref:XY_1D}, two~\cite{ref:XY_2D},
three~\cite{ref:XY_3D}, and four~\cite{ref:XY_4D} dimensions.
In particular, two-dimensional (2D) $XY$ model has been found to exhibit 
the famous Berezinskii-Kosterlitz-Thouless (BKT) transition~\cite{ref:XY_2D},
and the all-to-all globally-coupled $XY$ system has been known to exhibit the 
mean-field (MF) type phase transition~\cite{ref:XY_MF}. 
Also,  recent progress in complex network research has made 
the interplay between connection structure of interacting spins 
and the nature of phase transition a central topic
in statistical physics~\cite{ref:XY_complexnetwork,ref:XY_small}. 
Differently from most previous studies, we investigate in this paper
the one-dimensional (1D) $XY$ model with
variable interaction range and explore how the interaction range 
affects the collective critical behavior  of the model.  
The underlying 1D topology of the system allows us to study the
{\textit{winding number excitation}} that has been known to exist in the system
with local/nonlocal
interaction~\cite{ref:twiststate,ref:nonlocal_Hong_Kim}. 
We propose a novel nonlocal order parameter based on the distribution of the 
winding number excitation, which is found to successfully exhibit
a clear signature of the phase transition.

The present paper consists of five sections.  Section~II introduces the model
and in Sec.~III we explore thermodynamic behavior of the model, by measuring
the standard quantities such as magnetization, specific heat, susceptibility,
and Binder's cumulant to detect the phase transition. 
In Sec.~IV, which contains the key part of the present paper,
we define the winding number for a given configuration of the phase
variables, and investigate the behavior of its distribution depending on the
system size and temperature. Based on the observation of the winding
number distribution function, a novel nonlocal order parameter is introduced
and used to successfully detect the phase transition. 
Finally, a brief summary follows in Sec.~V. 
 
\section{Model}
We study the 1D $XY$ model of $N$ spins described by the Hamiltonian
\begin{equation}
\label{eq:XY_Hamiltonian}
H = - \frac{J}{2L}\sum_{i=1}^N \sum_{j=i-L}^{i+L}\cos(\phi_i - \phi_j),
\end{equation}
where $\phi_i \in (-\pi, \pi)$ is the phase angle variable
of the $i$th spin with the periodic boundary
condition $\phi_{i+N} = \phi_i$, and 
$L$ is the interaction range in one side so that each spin interacts with 
total $2L$ nearest-neighbor spins in both sides.
We take the ferromagnetic coupling with the positive strength $(J>0)$, which makes the neighbor spins favor 
their phase difference to be minimized.
The typical 1D $XY$ model with the nearest-neighbor 
interaction corresponds to the case of $L=1$, and the globally-coupled
$XY$ model with all-to-all couplings is achieved when $L=N/2-1$, respectively.
We in the present paper investigate the collective behavior of the model
given by Eq.~(\ref{eq:XY_Hamiltonian}), varying the interaction 
range $L$, which is a key control parameter 
that passes from the short-range ($L/N\rightarrow 0$) to the 
infinite-range  ($L/N \rightarrow 1/2$) regimes 
in the thermodynamic limit of $N\rightarrow \infty$. 

After suitable normalization of time, energy, and temperature,
the first-order Langevin-type equations of motion of the system read
\begin{eqnarray}  
\label{eq:dynamics}
\dot{\phi_i} &=& - \frac{\partial H}{\partial \phi_i} + \eta_i, \nonumber\\
             &=& - \frac{1}{2L} \sum_{j=i-L}^{i+L}\sin(\phi_i - \phi_j) + \eta_i, 
\end{eqnarray}  
where the dimensionless thermal noise $\eta_i$ satisfies
\begin{equation}
\label{eq:eta}
\langle \eta_i(t) \eta_j (t^{\prime})\rangle = 2 T \delta(t-t^{\prime})
\end{equation}
at the dimensionless temperature $T$ in units of $J/k_B$ with
the Boltzmann constant $k_B$.
We have taken the overdamped regime with the inertia term 
(containing $\ddot{\phi_i}$) neglected as in 
Ref.~\onlinecite{ref:PRB_35_8528} where a   
condensed matter system (smectic liquid-crystal)
has been studied. 
Equation~(\ref{eq:dynamics}) at 
zero temperature $(T=0)$ has also been used to describe 
the system of the 1D coupled oscillators 
with variable interaction range~\cite{ref:nonlocal_Hong_Kim}. 
If one is only interested in equilibrium behavior as in the present study,
one can use the Mont-Carlo (MC) simulation method instead. 

We note that the 1D system with short-range interaction does not exhibit any
long-range order at finite temperature~\cite{ref:XY_1D}.  On the other hand,
when the dimensionality of the system is higher than the lower critical dimension 2,
the celebrated Mermin-Wagner theorem~\cite{ref:mermin-wagner} cannot disprove
the existence of the long-range order, as typically exemplified
in 3D~\cite{ref:XY_3D}, 4D~\cite{ref:XY_4D}, and globally-coupled~\cite{ref:XY_MF} $XY$ models. 
In this context, it is natural to have the following questions: 
Is there a critical interaction range $(L/N)_c$ beyond which a finite-temperature
phase transition starts to occur?  If so, does the phase transition always belong to the
MF universality class?
To answer these questions, we study the critical behavior of the 
model, first measuring some standard quantities for various values of 
$L$ and $N$, which is presented in next section.

\section{Thermodynamic Behavior}
In this section we numerically study the thermodynamic behavior of the system,
with particular attention paid to the emergence of the phase transition at
finite values of the interaction range $L$.  It is possible to numerically
investigate equilibrium behavior of the system in two different ways: 
numerical integration of
equations of motion Eq.~(\ref{eq:dynamics}) and MC simulation based
on the Hamiltonian Eq.~(\ref{eq:XY_Hamiltonian}). It is straightforward to show
that both numerical methods are equivalent to each other as long as only
equilibrium behaviors are concerned: the steady-state solution of the
Fokker-Planck equation based on Eq.~(\ref{eq:dynamics}) is simply the Boltzmann
distribution $P \sim \exp(-H/k_BT)$ with the Hamiltonian in
Eq.~(\ref{eq:XY_Hamiltonian})~\cite{Fokker_Planck}.  
Since the MC simulations usually run much faster than direct
integration of stochastic differential equations, we 
here use the former with the standard Metropolis 
local update algorithm and measure various quantities of interest. 
In the MC simulations, all quantities are measured over 
$10^7$-$10^8$ MC steps after equilibration over 
the initial $10^6$ MC steps.

\begin{figure}
\epsfxsize=1.0 \linewidth \epsfbox{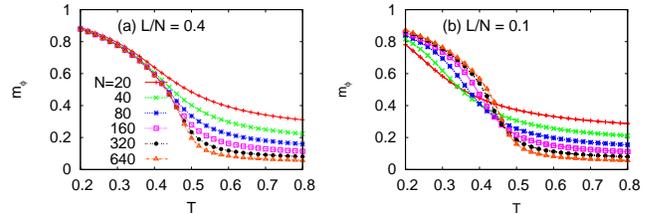}
\caption{(Color online) Magnetization $m_{\phi}$ is plotted 
as a function of the temperature $T$ 
for various system sizes $N$ and interaction ranges $L$ 
for (a) $L/N=0.4$ and (b) $L/N=0.1$, respectively. 
} 
\label{fig:m}
\end{figure}

We measure the equilibrium magnetization defined as 
\begin{equation} 
\label{eq:m}
m_{\phi} = \left\langle \left| \frac{1}{N}\sum_{j=1}^N e^{i\phi_j} \right| \right\rangle, 
\end{equation}
where $\langle \cdots \rangle$ denotes the thermal average. 
The behavior of the magnetization $m_{\phi}$ is shown as a function of the 
temperature $T$, varying the value of $L$ to keep
the ratio $L/N$ unchanged~[see Fig.~\ref{fig:m}]. 
We find that the magnetic ordering ($m_{\phi} > 0$) occurs at $T \lesssim 0.5$ for $L/N=0.4$, as 
shown in Fig.~\ref{fig:m}(a). 
The signature of the transition from the disordered phase ($m_{\phi}=0$) to 
the ordered one ($m_{\phi}>0$) becomes clearer as the system size $N$ increases. 
We also measure the magnetization for smaller value, 
$L/N=0.1$ [see Fig.~\ref{fig:m}(b)] and $L/N=0.05$ (not shown here), and 
find that the two cases also show the magnetic ordering for $T\lesssim 0.5$,
for sufficiently large system sizes. 
The behavior of the magnetization shown in Fig.~\ref{fig:m} suggests 
that the finite-temperature phase transition occurs at any nonzero 
value of $L/N$. In other words, we expect that the critical
value of the interaction range beyond which the phase transition
occurs at a nonzero critical temperature satisfies $(L/N)_c=0$ 
in the thermodynamic limit.

\begin{figure}
\epsfxsize=1.00 \linewidth \epsfbox{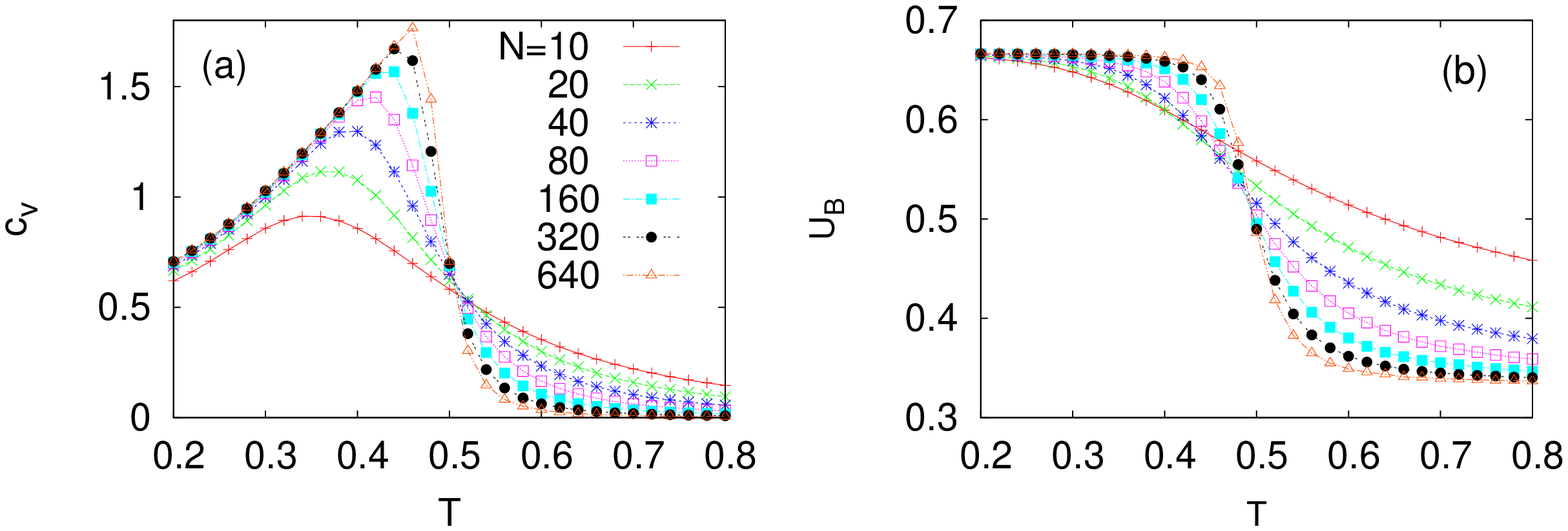}
\caption{(Color online) (a) Specific heat $c_v$
and (b) Binder's cumulant $U_B$ versus
the temperature $T$ for various system sizes $N$. 
The interaction rage $L$ is chosen
to yield the same value of the ratio $L/N=0.4$.  
} 
\label{fig:all4}
\end{figure}

\begin{figure}
\epsfxsize=1.00 \linewidth \epsfbox{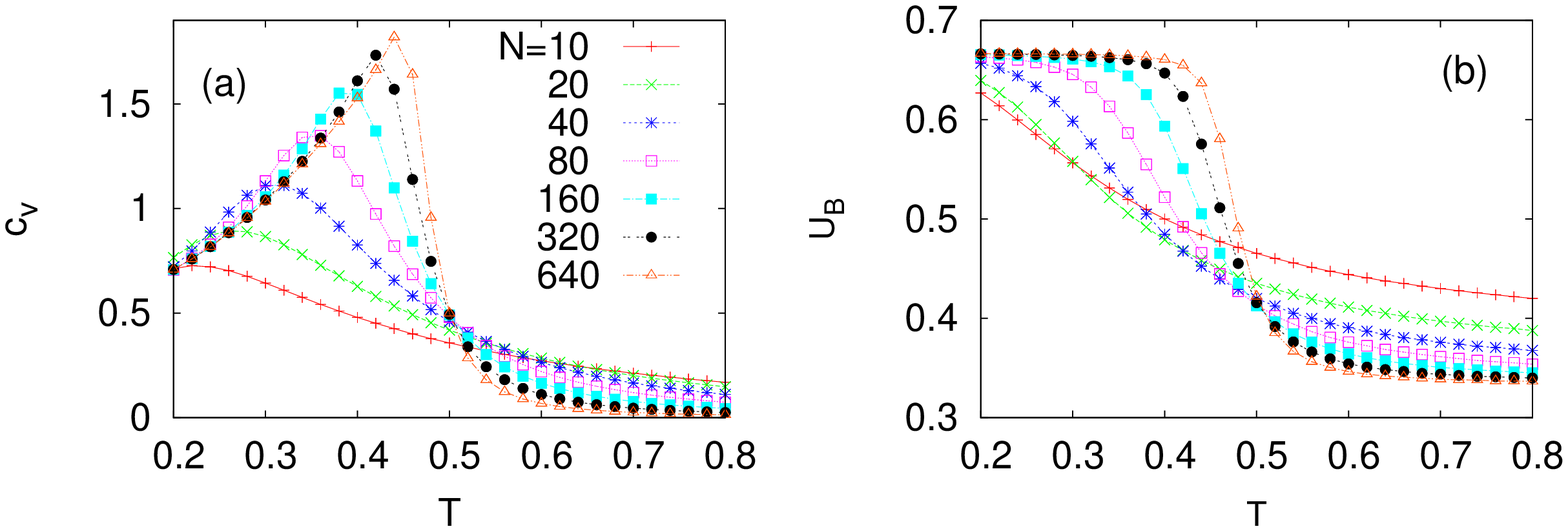}
\caption{(Color online) (a) $c_v$ and (b) $U_B$ versus $T$ 
for $L/N=0.1$. For comparison, see Fig.~\ref{fig:all4} for $L/N=0.4$.}
\label{fig:all1}
\end{figure}

To see it further, we also investigate other standard quantities such as the 
specific heat $c_v$, susceptibility $\chi$, and Binder's fourth-order cumulant 
$U_B$~\cite{ref:Binder},  defined by
\begin{eqnarray} 
c_v &\equiv&  \frac{1}{N k_B T^2}\Bigg( \langle H^2 \rangle - \langle H \rangle^2 \Bigg), \label{eq:Cv}\\
\chi &\equiv& \frac{N}{k_B T} \Bigg(\langle m^2_{\phi} \rangle - \langle m_{\phi} \rangle^2 \Bigg), \label{eq:chi} \\
U_B &\equiv& 1 - \frac{ \langle m^4_{\phi} \rangle } { 3 \langle m^2_{\phi} \rangle^2 }, \label{eq:U}
\end{eqnarray}
where $H$ is the  Hamiltonian in Eq.~(\ref{eq:XY_Hamiltonian}).
Figures~\ref{fig:all4} and \ref{fig:all1} show $c_v$ and $U_B$
as functions of $T$ for $L/N=0.4$ and $L/N=0.1$, respectively. 
We find that both quantities, $c_v$ and $U_B$, strongly 
support the emergence of the phase transition at $T \approx 0.5$ for both 
$L/N=0.4$ and $0.1$. Although not shown here $\chi(T)$ has a peak which shifts towards $T_c \approx 0.5$
as the system size is increased both for $L/N = 0.4$ and 0.1.
We also investigated for the case of $L/N=0.05$ (not shown), and obtained 
the same conclusion if we focus on larger system sizes.

We turn our attention to the universality class of the 
phase transition, and consider the critical behavior of the magnetization 
characterized by 
\begin{equation}
\label{eq:m_infiniteN}
m_{\phi} \sim (T - T_c)^\beta
\end{equation}
with the critical exponent $\beta$ and the critical temperature $T_c$ in the thermodynamic 
limit. 
According to the finite-size scaling theory~\cite{ref:FSS}, 
we expect that 
$m_{\phi}$ in a finite-sized system satisfies 
the scaling form 
\begin{equation} 
\label{eq:mphi_scaling}
m_{\phi} = N^{-\beta/\bar\nu} f\bigl( (T{-}T_c) N^{1/{\bar\nu}} \bigr),
\end{equation}
where $\bar\nu$ is the critical exponent that describes the critical 
behavior of the correlation volume $\xi_v \sim \xi^d$ in $d$ dimensions: 
$\xi_v \sim |T-T_c|^{-\bar\nu}$~\cite{ref:xi_v}.
The scaling function $f(x)$ with the scaling variable $x \equiv (T-T_c)N^{1/\bar\nu}$
has limiting behaviors:
$f(x) \sim x^{\beta}$ as $x \rightarrow 0$ and $f(x)\sim {\rm{const.}}$ as 
$x \rightarrow +\infty$.  
At criticality ($T=T_c$), the magnetization reduces to 
\begin{equation}
\label{eq:m_Tc}
m_{\phi} \sim N^{-\beta/\bar\nu}.
\end{equation}

\begin{figure}
\epsfxsize=1.0 \linewidth \epsfbox{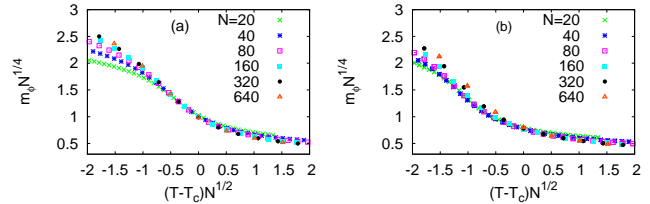}
\caption{(Color online) Finite-size scaling collapse of the magnetization $m_{\phi}$: $m_{\phi}N^{\beta/\bar\nu}$ versus $(T-T_c)N^{1/\bar\nu}$ 
for (a) $L/N=0.4$ and (b) $L/N=0.1$. With $T_c=0.5$, $\beta/\bar\nu=1/4$, and $\bar\nu=2$ chosen,
good quality of scaling collapse of data is achieved.} 
\label{fig:mscale}
\end{figure}

To estimate the exponents $\beta$ and $\bar\nu$ we 
plot $m_{\phi}$ as a function of the size $N$ in the log-log scale (not shown), and measure 
its slope, which gives us the exponent $\beta/\bar\nu$.   
We also check the finite-size scaling 
relation directly by plotting $m_{\phi}N^{\beta/\bar\nu}$ versus 
$(T-T_c)N^{1/\bar\nu}$, controlling the values of $\beta/\bar\nu$ and 
$\bar\nu$ [see Fig.~\ref{fig:mscale}(a)].
We find that the scaling function $f(x)$ converges to a constant 
for large $x$, and behaves as $x^{\beta}$ for small $x$ as expected.
The numerical findings shown in Fig.~\ref{fig:mscale}(a) is then summarized as
\begin{equation}
\beta/\bar\nu=1/4~~{\mbox{and}}~~\bar\nu=2,
\end{equation}
which yields $\beta=1/2$. 
This result implies that the nature of the phase transition for $L/N=0.4$ is 
the same as that of the MF transition~\cite{ref:XY_MF}. The crossing of the
specific heat $c_v$ at $T_c$ shown in Fig.~\ref{fig:all4}(a) and Fig.~\ref{fig:all1}(a)
also implies that the specific heat exponent $\alpha = 0$, in accord with the MF 
universality class~\cite{ref:XY_small}. 
We also examined the case for the smaller 
value of $L/N=0.1$ [see Fig.~\ref{fig:mscale}(b)] and $L/N=0.05$ (not shown), and 
obtained the same result, which suggests that the mean-field nature
of phase transitions should be robust for any nonzero value of $L/N$.

\section{Winding number excitation and a nonlocal order parameter}
Recently, the emergence of {\textit{twisted wave}} in the 
system of coupled oscillators with local/nonlocal 
interaction has been 
reported~\cite{ref:twiststate,ref:nonlocal_Hong_Kim}. 
We note that the development of the twisted wave in dynamic models 
has the same physical origin as the {\textit{winding number excitation}} in
the 1D $XY$ model. 
In this section, we measure the winding number across the system in equilibrium
for each sample and compute its probability distribution function by using
a large-size ensemble of samples. 
We define the winding number $q$ for a given phase configuration by 
\begin{equation}
\label{eq:q}
q \equiv \frac{1}{2\pi}\sum_{i=1}^N {\mbox{mod}}(\phi_{i+1} - \phi_i), 
\end{equation}
where `mod' denotes that the phase difference $\phi_{i+1} - \phi_i$ is 
measured modulo $2\pi$ and the periodic boundary condition $\phi_{N+1} = \phi_1$
is used.  
It is to be noted that the gapless spin-wave-type excitation can occur 
without changing the winding number. This reminds us the   
topological excitation of vortices in the conventional 
2D $XY$ model~\cite{ref:XYreview,ref:XY_2D}. 
\begin{figure}
\epsfxsize=1.0 \linewidth \epsfbox{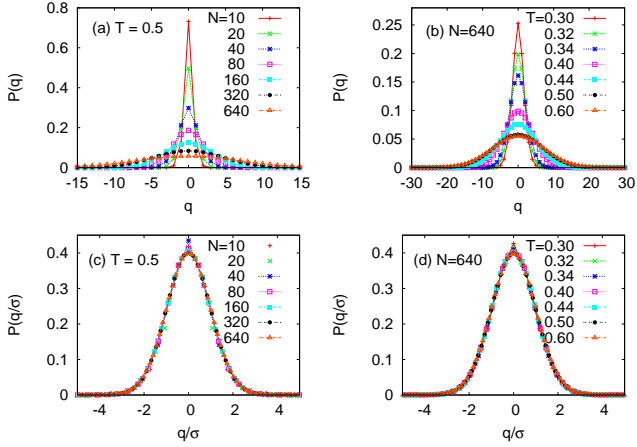}
\caption{(Color online) Probability distribution function $P(q)$ of the winding
number $q$ (a) at $T=0.5$ for various system sizes $N$ and (b) for a given size $N=640$
at various temperatures $T$. For each $P(q)$ the width of the distribution $\sigma(N,L,T)$
is computed and then used to scale the horizontal axis as shown in (c) and (d), each
obtained from (a) and (b), respectively. For (a)-(d), $L/N=0.4$ has been used. 
}
\label{fig:pq4}
\end{figure}

From the observation of the standard thermodynamic quantities in Sec.~III, 
the system surely undergoes a single phase transition. Accordingly, we expect
that the existence of the phase transition should also alter the 
pattern of the winding number excitation in some way
at the observed critical temperature
$T_c \approx 0.5$. 
We measure the probability distribution function of the 
winding numbers to see a signature of the phase transition, after 
sufficient equilibration procedure.
Figure~\ref{fig:pq4}(a) shows the distribution $P(q)$ for various system 
sizes $N$ at the critical temperature $T_c(=0.5)$. 
We find that the width of the distribution function $P(q)$ systematically
changes, as the system size $N$ is increased. The inversion symmetry 
$P(-q) = P(q)$ is easily understood since the system has no reason to prefer
clockwise winding to counterclockwise one (and vice versa).
The temperature dependence of the distribution $P(q)$ is also investigated at a given system 
size $N=640$ [see Fig.~\ref{fig:pq4}(b)]. Again, we observe that the width of the distribution
changes as $T$ is varied. In order to check the possibility of the scaling of the winding
number distribution function, we measure the standard deviation $\sigma$ defined by 
\begin{equation}
\label{eq:sigma}
\sigma \equiv \sqrt{\langle q^2 \rangle - \langle q \rangle^2}, 
\end{equation}
where $\langle q^2 \rangle = \int q^2 P(q) dq$ 
and $\langle q \rangle = \int q P(q) dq$.
We then scale the horizontal axis in Fig.~\ref{fig:pq4}(a) and (b)
by using the scaling variable $q/\sigma$ with $\sigma$ numerically
computed for given values of $N$, $L$, and $T$. 
It is shown very clearly that after the winding number $q$ is scaled by
the width $\sigma$ of the distribution function, all the curves are put
on top of each other as displayed in Fig.~\ref{fig:pq4}(c) and (d), 
obtained from Fig.~\ref{fig:pq4}(a) and (b), respectively.
Our observation of the collapse of the winding number distributions
strongly suggests that the width  $\sigma(N,L,T)$ of the distribution function
can successfully represent the whole distribution function and thus will be used
below for further analysis to study critical behavior in the system.

\begin{figure}
\epsfxsize=1.0 \linewidth \epsfbox{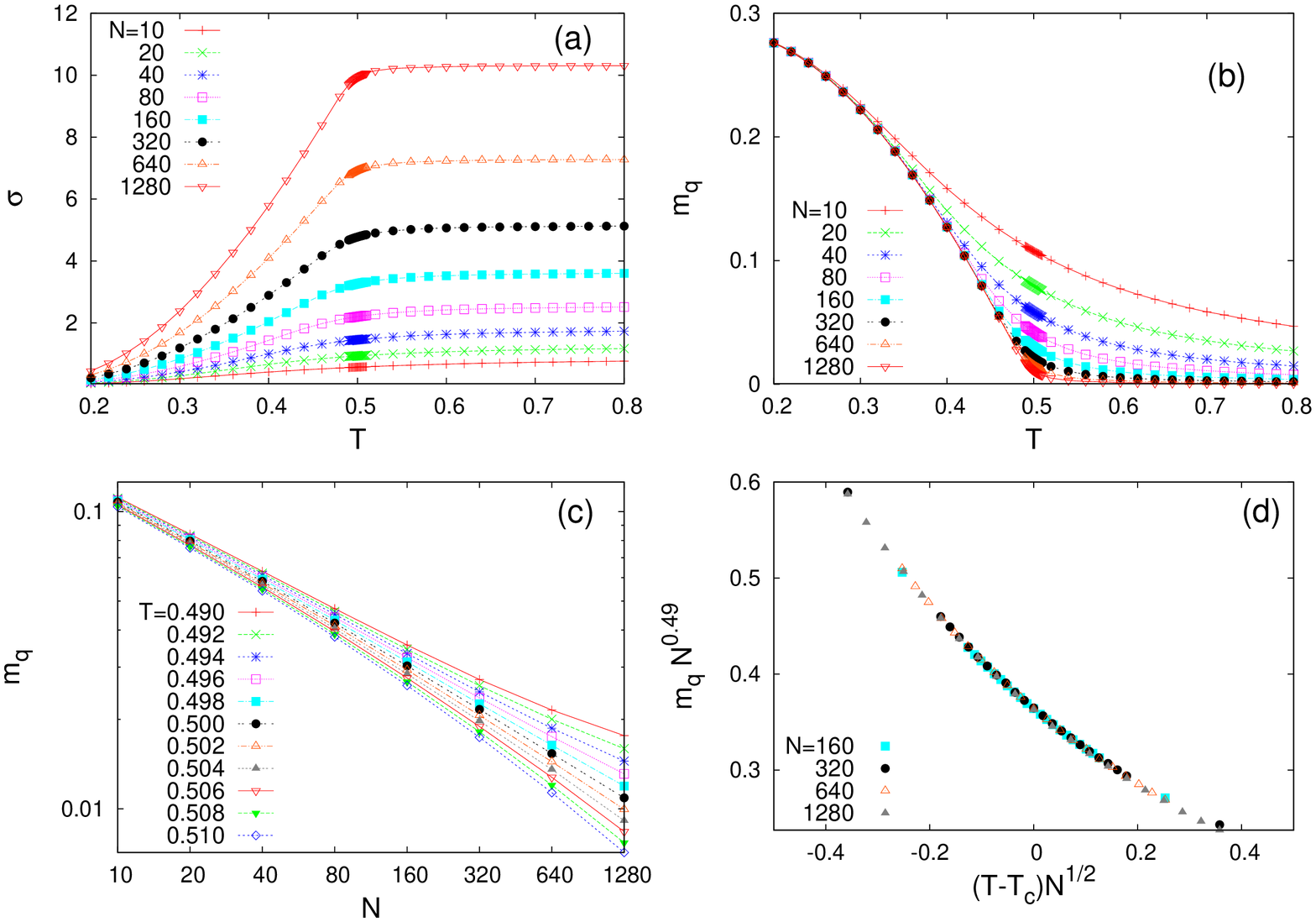}
\caption{(Color online) 
(a) Standard deviation $\sigma$ is plotted as a function of the 
temperature $T$ for various system sizes $N$. (b) The newly defined nonlocal order parameter 
$m_q$ in Eq.~(\ref{eq:mq}) versus $T$ for various values of $N$. 
(c) $m_q$ versus $N$ in log-log scale clearly exhibits the power-law decay
form at $T=T_c \approx 0.5$, with the slope corresponding to
$\beta_q/\bar\nu_q=0.49$. 
(d) Finite-size scaling collapse of $m_q$ yields $\bar\nu_q=2.00(5)$. 
From (c) and (d), we estimate $T_c = 0.500(3)$,  $\beta_q/\bar\nu_q = 0.49(4)$
and $\bar\nu_q = 2.00(5)$. 
All results in (a)-(d) are for $L/N = 0.4$.
} 
\label{fig:sigma4}
\end{figure}

In Fig.~\ref{fig:sigma4}(a) we show $\sigma$ as a function of $T$ for various
system sizes with $L/N = 0.4$ fixed.  From the central-limit theorem
for the independent random variables, we expect that 
$P(q)$ at infinite temperature has $\sigma \sim \sqrt{N}$. 
In order to compensate such a behavior, we define
the normalized width $s \equiv \sigma/\sqrt{N}$, and 
introduce a new quantity $m_q$ defined by
\begin{equation}
\label{eq:mq} 
m_q \equiv s_{\infty} - s,
\end{equation}
where $s_{\infty} \equiv s(N \rightarrow \infty, T \rightarrow \infty)$. 
From Fig.~\ref{fig:sigma4}(a), it is expected that
$m_q \rightarrow 0$ from above in the high-temperature limit and
$m_q > 0$ in the low-temperature regime, hopefully playing the role of
the order parameter.
The calculation of the value of $s_{\infty}$ is straightforward
since the phase difference of the nearest neighbors $\phi_{i+1} - \phi_i$ 
simply becomes an independent random variable due to the lack of any spatial correlation
in the high-temperature limit. In other words, $x_i \equiv \mod(\phi_{i+1} - \phi_i)$ 
is independent from each other and randomly distributed in $x_i \in (-\pi, \pi]$.
From the symmetry of the distribution we get $\langle q \rangle = 0$, 
and the second moment $\langle q^2 \rangle$ is computed as
\begin{eqnarray} 
\label{eq:q2}
\langle q^2 \rangle &=& \frac{1}{4\pi^2} \Bigg\langle 
\Bigg(\sum_{i=1}^{N} x_i \Bigg)^2 \Bigg\rangle \nonumber\\
 &=& \frac{N}{4\pi^2} \int_{-\pi}^{\pi} P(x) x^2 dx = \frac{N}{12},
\end{eqnarray} 
where $P(x)$ is the uniform probability distribution function for $x$ and
the cross terms $x_i x_j~(i\neq j)$ has given null contribution from the independence between 
$x_i$ and $x_j$ at infinite temperature. 
Consequently, we get $\sigma = \sqrt{\langle q^2 \rangle -\langle q \rangle^2} = \sqrt{N}/(2\sqrt{3})$
in the high-temperature limit, 
yielding $s_{\infty} = \sigma/\sqrt{N} = 1/(2\sqrt{3}) \approx 0.288675$. We then 
plot Fig.~\ref{fig:sigma4}(b) for $m_q$ versus $T$, which looks very much similar to the standard
order parameter $m_\phi$ in Fig.~\ref{fig:m}. 
Finite-size scaling ansatz then allows us to expect the existence of the scaling function given by
\begin{equation} 
\label{eq:mq_scaling}
m_q = N^{-{\beta_q}/\bar\nu_q} g \bigl( (T - T_c) N^{1/\bar\nu_q} \bigr),
\end{equation}
where $g(x)$ is a scaling function that behaves as $g(x) \sim x^{\beta_q}$ as 
$x\rightarrow 0$, and $g(x)\sim {\mbox{const.}}$ as $x\rightarrow +\infty$. 
At criticality $(T=T_c)$, the new nonlocal order parameter $m_q$ is expected to 
show the power-law behavior: $m_q \sim N^{-\beta_q/\bar\nu_q}$.
On the basis of the prediction, we detect the exponents 
$\beta_q$ and $\bar\nu_q$, by plotting $m_q$ as a function of 
$N$ for various $T$, where the slope at $T_c$ gives the value of 
$\beta_q/\bar\nu_q$~[see Fig.~\ref{fig:sigma4}(c)].
If only the large system sizes ($N \geq 160$) are used, we find that
the curves follow the power-law form in Fig.~\ref{fig:sigma4}(c)
at $T = 0.497$-$0.502$ with the slopes $\beta_q/\bar\nu_q = 0.45$-$0.52$.
For temperatures outside of this range, the curves exhibit clear deviations
from the power-law form.
The best fit to the power-law form is obtained at $T = 0.500$ and $\beta_q/\bar\nu_q = 0.49$,
and thus we conclude $T_c = 0.500(3)$ and $\beta_q/\bar\nu_q = 0.49(4)$.
The exponent $\bar\nu_q$ is obtained from the data collapse: 
Figure~\ref{fig:sigma4} (d) shows the behavior of $m_q N^{\beta_q/\bar\nu_q}$ against 
$(T-T_c) N^{1/\bar\nu_q}$, where $\beta_q/\bar\nu_q=0.49$ and 
$\bar\nu_q=2.00$ at $T_c=0.500$ are 
used for the best collapse,  yielding $\beta_q=0.98(11)$. 
It is particularly important to recognize that our new nonlocal order parameter
based on the topological winding number excitation gives the correlation exponent
$\bar\nu_q = 2$, identical to $\bar\nu = 2$ previously confirmed for the standard
local magnetization order parameter $m_\phi$.
However, we believe that further study is required to understand the value of 
$\bar\beta_q \approx 0.98$ which is very close to unity.
We also examine results for $L/N=0.1$ and $L/N=0.05$ (not shown here); although not as
clear as for $L/N=0.4$, we are able to confirm the same values of exponents
as long as sufficiently bigger system sizes are used.

\section{Summary}
In summary, we have numerically investigated the critical
behavior of  the 1D $XY$ model of $N$ spins with variable interaction range $L$. 
It has been confirmed that the critical interaction range beyond
which the phase transition starts to occur at a nonzero finite temperature is very
small, and presumably $(L/N)_c = 0$. 
The nature of the transition has been examined by measuring 
standard quantities such as the magnetization, specific heat, susceptibility, 
and Binder cumulant. All measured quantities unanimously suggest the transition
is of the mean-field type at any nonzero value of $L/N$. 
The underlying one-dimensional topology of the system makes it possible to 
study the winding number excitation. By systematically examining the
probability distribution function of the winding number excitation, we have
suggested a novel nonlocal order parameter $m_q$ based on the width of the winding
number distribution function. 
We show that our new order parameter $m_q$ can successfully detect the phase
transition and its nature.

\section*{Acknowledgment}
This work was supported by Basic Science Research Program through 
Grant No. NRF-2012R1A1A2003678 (H.H.) and No. NRF-2014R1A2A2A01004919 (B.J.K.).

\end{document}